\documentclass[showpacs,preprintnumbers,amsmath,amssymb,aps]{revtex4}
\def\unit{\leavevmode\hbox{\small1\kern-3.6pt\normalsize1}}
 \normalsize

\def\lsim{\raise0.3ex\hbox{$\;<$\kern-0.75em\raise-1.1ex\hbox{$\sim\;$}}}
\def\gsim{\raise0.3ex\hbox{$\;>$\kern-0.75em\raise-1.1ex\hbox{$\sim\;$}}}
\newcommand{\x}[2]{#1 \times #2}
\newcommand{\mat}[1]{\begin{pmatrix} #1 \end{pmatrix}}

\usepackage{graphics}

\newcommand{\be}{\begin{eqnarray}}
\newcommand{\ee}{\end{eqnarray}}

\newcommand{\n}{\nonumber\\}
\newcommand{\nn}{\tilde{\nu}}

\def\bea{\begin{eqnarray}}
\def\eea{\end{eqnarray}}

\usepackage{epsfig}
 \normalsize

\begin{document}
\title{Right-handed sneutrino-antisneutrino oscillations\\
in a TeV scale Supersymmetric $B-L$ model}
\author{A. Elsayed$^{1,2}$, S. Khalil$^{1,3,4}$, S. Moretti$^{4,5}$, A. Moursy$^{1}$ }
\vspace*{0.2cm}
\affiliation{$^1$Center for Theoretical Physics, Zewail City for Science and Technology, 6 October City, Cairo, Egypt.\\
$^2$Department of Mathematics, Faculty of Science, Cairo
University, Giza, 12613, Egypt.\\
$^3$Department of Mathematics, Faculty of
Science,  Ain Shams University, Cairo, 11566, Egypt.\\
$^4$School of Physics and Astronomy, University of Southampton,
Highfield, Southampton SO17 1BJ, UK.\\
$^5$Particle Physics Department, Rutherford Appleton Laboratory,
Chilton, Didcot, Oxon OX11 0QX, UK.}
\date{\today}

\begin{abstract}
We explore right-handed sneutrino-antisneutrino mixing in a TeV
scale $B-L$ extension of the Minimal Supersymmetric Standard Model (MSSM),
$(B-L)$SSM, where a type I seesaw mechanism of light neutrino mass
generation is naturally implemented.  The constraints
imposed on the mass splitting between heavy right-handed sneutrino and
the corresponding antisneutrino by the experimental limits set on the
light neutrino masses are investigated. We also study direct pair
production of such right-handed sneutrinos at the Large Hadron Collider (LHC) and its decay
modes, emphasising that their decay into same-sign di-lepton pairs
are salient features for probing these particles at the CERN machine. Finally,
the charge asymmetry present in such same-sign di-lepton signals is also analysed and
confirms itself as a further useful handle to extract information about the oscillation dynamics.
\end{abstract}
\maketitle

\section{Introduction}
\label{sec:intro}
Experimental evidence exists for the oscillation into one another of physical eigenstates of relativistic fields/particles with degenerate quantum numbers, specifically, in neutral systems, like $K^0$, $B^0_d$, $B^0_s$, $D^0$, and, most importantly for our study, $\nu$'s. In the case of the hadronic states, the associated measurements provided important insights  into the structure of Electro-Weak (EW) interactions, in particular, their Charge and Parity (CP) dynamics, confirming that CP-violation indeed occurs in Nature. In the case of neutrinos, proof that they oscillate translates into the fact that they have non-zero masses. Hence, no matter where such the oscillation phenomenon appears, it has always
lead to
clear advances in the understanding of the fundamental interactions governing the behaviour of fields and particles.

Conversely, from a theoretical point of view, we know that the Standard Model (SM), whereas it can account for oscillations in the aforementioned hadronic systems, cannot explain neutrino masses, as the latter are, by construction, absent in it.
Hence, some Beyond the SM (BSM) physics ought to be invoked to accommodate neutrino oscillations. Further, if one
recalls the so-called hierarchy problem of the SM, its inability to provide a candidate for Dark Matter (DM), its
failure to explain the matter-antimatter asymmetry in the Universe and the lack of gauge coupling unification at any scale
in it, a natural way forward in  the quest to formulate a viable BSM scenario is to adopt
Supersymmetry (SUSY), which can remedy at once all such flaws. In fact, SUSY can also easily co-exists with constraints emerging from the hadronic
and leptonic sectors, when it comes to incorporate oscillation phenomena.

The simplest realisation of SUSY is the Minimal Supersymmetric Standard Model (MSSM), whereby
the gauge structure of
the SM is maintained and the matter spectrum is limited to the most economical structure able to ensure anomaly cancellations, which corresponds to the adoption of an additional Higgs doublet field (with respect to the SM) and the
Supersymmetrisation of all ensuing Higgs/gauge boson and  fermion fields into the Higgsinos/gauginos and sfermions.
The MSSM is therefore rather predictive amongst the many possible SUSY realisations, in so far that the number of independent
parameters defining it (whether it be in a constrained or unconstrained version) is always the smallest
with respect to those of its
SUSY alternatives. However, this feature also renders it very testable from an experimental point of view. From this perspective, it is now widely acknowledged that Large Hadron Collider (LHC) data have reduced considerably the viable
parameter space of the MSSM, in both its constrained and unconstrained formulation, and/or confined it to rather unnatural
spectrum configurations (of masses and couplings). On the one hand, the very recent observation of a SM-like Higgs boson constrains significantly the spectrum of the MSSM Higgs sector, namely, $\tan\beta$ (the ratio of the Vacuum Expectation
Values (VEVs) of the two aforementioned Higgs doublet fields) and the mass of one of the Higgs bosons
(there are five in the MSSM, three neutral and a pair of charged ones) directly as well as the mass of the squarks
 (the SUSY counterparts of the SM quarks) indirectly.  On the other hand, no observation whatsoever of the aforementioned Higgsinos, gauginos and sparticles at the mass scale
preferred by the MSSM to achieve gauge coupling unification (i.e., of order 1 TeV or so) affirms the possibility that the SUSY  sparticle sector is richer than the one of the MSSM.

In the light of all this, it has therefore become of extreme relevance to explore non-minimal realisations of SUSY, better
compatible with current data than the MSSM yet similarly predictive. Owing to the established presence of non-zero neutrino masses, a well motivated path to follow in this direction is to consider the $B-L$ Supersymmetric Standard Model, henceforth
$(B-L)$SSM for short. Herein,  (heavy) right-handed neutrino
Superfields are introduced in order to implement a type I
seesaw mechanism, which provides an elegant solution for the existence and
smallness of the (light) left-handed neutrino masses. Right-handed neutrinos can be
naturally implemented in the $(B-L)$SSM, which is based on the gauge group $SU(3)_C \times
SU(2)_L \times U(1)_Y \times U(1)_{B-L}$, hence the
simplest generalisation of the SM gauge group (through an additional $U(1)_{B-L}$ symmetry). In this model, it has been shown that the scale of $B-L$ symmetry breaking is related to the SUSY breaking
scale \cite{Khalil:2007dr}, so that this SUSY realisation predicts several testable signals at the LHC,
not only in the sparticle domain but also in the $Z'$  (a $Z'$ boson in fact emerges from the $U(1)_{B-L}$
breaking), Higgs (an additional singlet state is economically introduced here, breaking the $U(1)_{B-L}$ group) and (s)neutrino sectors \cite{Khalil:2006yi,B-L-LHC,PublicPaper}. Furthermore, other than assuring its testability
at the LHC, in fact in a richer form than the MSSM, the $(B-L)$SSM also alleviates the so-called `little' hierarchy problem
of the MSSM, as both the additional singlet Higgs state and right-handed (s)neutrinos
\cite{BLMSSM-Higgs,O'Leary:2011yq,Basso:2012ew}
release additional parameter space from the LEP, Tevatron and LHC bounds.
A DM candidate plausibly different from the MSSM one exists as well \cite{Basso:2012gz}.
Finally, interesting results on the ability of the $(B-L)$SSM to emulate the Higgs boson signals isolated at the LHC
are also emerging \cite{BLMSSM-LHC}.

Now, if the heavy (anti)neutrinos are of the Majorana type, as it is the case for the $(B-L)$SSM, it is expected
that their SUSY partners, i.e., the (anti)sneutrinos, oscillate
analogously to the physics systems recalled before. That is,
sneutrino-antisneutrino mixing occurs and the $\Delta L = 2$ Majorana neutrino mass terms governing such a dynamics
can induce a mass splitting ($\Delta m_{\tilde \nu}$) between the physical states. The effect of this mass splitting is to induce sneutrino-antisneutrino oscillations {\cite{hirschetal,grossman-haber1}}. This can in turn lead to a sneutrino decaying into a final state with a ``wrong-sign charged lepton''.
If such sneutrino-antisneutrino system is produced via pairs at the LHC, the ultimate result is that
 same-sign di-lepton pairs are eventually induced at the LHC, that can then be studied to reveal such an oscillation.
In fact, the charge asymmetry of such same-sign di-lepton signals produced
by right-handed sneutrino-antisneutrino oscillations can also be analysed at the CERN machine,
providing further insights into such a phenomenon.

Much of this phenomenology has been tackled already in the MSSM, see Ref.~\cite{KhalilLFV,oscillations-MSSM}
(see also \cite{Chun:2001mm} for the $R$-parity violating case). It is the purpose
of this paper, motivated by the previous discussion about the drawbacks of the MSSM, to re-address the analysis
to the case of the $(B-L)$SSM. The paper is organised as follows. In the next section we describe the theoretical setup of
the right-handed (anti)sneutrino sector in such a model while Sect.~\ref{sec:constraints} discusses the experimental
constraints on it.  Sect. \ref{sec:results} will instead present our results for the LHC. Finally, we conclude
in Sect.~\ref{sec:summa}.

\section{Theory of right-handed sneutrino-antisneutrino mixing}
\label{sec:sneu}
In this section, we analyse the right-handed sneutrino sector in the $(B-L)$SSM. As advocated
in the introduction, this type of extension implies the existence
of three extra Superfields, one per generation, with $B-L$ charge$=-1/2$, in order to cancel the associated $B-L$ triangle anomaly.
These Superfields are identified with the right-handed neutrinos
and will be denoted by $N_i$. In addition, in order to break the
$B-L$ symmetry at the TeV scale, two Higgs Superfields
$\hat{\chi}_{1,2}$, with $\mp 1$ $B-L$ charges, are required. In
Table \ref{particle-content} we present the particle contents of the $(B-L)$SSM as well as the quantum
numbers of the chiral Superfields with respect to
$\x{SU(3)_C}{\x{SU(2)_L}{\x{U(1)_Y}{U(1)_{B-L}}}}$.\\

\begin{table}[th]
\begin{center}
\begin{tabular}{||c||c|c|c|c|c|c|c|c|c|c||}\hline\hline
   & $\hat{Q}_i$ & $\hat{U}_i^c$ & $\hat{D}_i^c$ & $\hat{L}_i$ & $\hat{E}_i^c$ & $\hat{N}_i^c$ &  $\hat{H}_1$ & $\hat{H}_2$ & $\hat{\chi}_1$ & $\hat{\chi}_2$\\
\hline\hline
$SU(3)_c$ & $3$ & $\bar{3}$ & $\bar{3}$ & $1$ & $1$ & $1$ & $1$ & $1$ & $1$ & $1$\\
\hline
$SU(2)_L$ & $2$ & $1$ & $1$ & $2$ & $1$ & $1$ & $2$ & $2$ & $1$ & $1$\\
\hline
$U(1)_Y$ & $1/6$ & $-2/3$ & $1/3$ & $-1/2$ & $1$ & $0$ & $-1/2$ & $1/2$ & $0$ & $0$\\
\hline
$U(1)_{B-L}$ & $1/6$ & $-1/6$ & $-1/6$ & $-1/2$ & $1/2$ & $1/2$ & $0$ & $0$ & $-1$ & $1$\\
\hline\hline
\end{tabular}
\caption{Chiral Superfields of the $(B-L)$SSM and their quantum
numbers}
\label{particle-content}
\end{center}
\end{table}

The $(B-L)$SSM Superpotential is given by the expression
\begin{equation}
{\cal W} = Y_u \hat{Q} \hat{H}_2 \hat{U}^c + Y_d \hat{Q} \hat{H}_1
\hat{D}^c  + Y_e \hat{L} \hat{H}_1 \hat{E}^c + Y_{\nu} \hat{L}
\hat{H}_2 \hat{N}^c + \frac{1}{2} Y_N \hat{N}^c \hat{\chi}_1
\hat{N}^c + \mu \hat{H}_1 \hat{H}_2 + \mu' \hat{\chi}_1
\hat{\chi}_2 .
\end{equation}
 The relevant soft SUSY breaking terms are given by
\begin{eqnarray}
- {\cal L}_{soft} &=&-{\cal L}^{\rm MSSM}_{soft} +
{\widetilde{m}}_{Nij}^{2}{\widetilde{N}}_{i}^{c*
}{\widetilde{N}}_{j}^{c} + m^2_{\chi_1} \vert{\chi_1}\vert^2 +
 m^2_{\chi_2}\vert{\chi_2}\vert^2 \nonumber\\
 &+& \left[ Y_{\nu ij}^{A}{\widetilde{L}}_{i}
{\widetilde{N}^c}_{j}H_{u} + Y_{N ij}^{A}{\widetilde{N}}_i^{c}
{\widetilde{N}}_j^{c}\chi_{1} + B \mu^\prime \chi_1 \chi_2 +
\frac{1}{2} M_{B-L}{\widetilde{Z}_{B-L}}{\widetilde{Z}_{B-L}}+
M'_{B-L}{\widetilde{Z}}{\widetilde{Z}_{B-L}} + h.c
\right] ,~~ %
\label{Lsoft}%
\end{eqnarray}%
where $(Y_N^A)_{ij}\equiv (Y_N A_N)_{ij}$ is the trilinear
associated with the Majorana neutrino Yukawa coupling. Note that, due
to the $B-L$ invariance, the bilinear coupling $B_N
\tilde{N^c}\tilde{N^c}$ is not allowed. It may be generated only
after the $B-L$ symmetry breaking. The Majorana mass $M_{B-L}'$ is
generated due to the possible gauge kinetic mixing between the two
Abelian groups $U(1)_Y$ and $U(1)_{B-L}$. It is known that this
gauge mixing term may be absorbed in the covariant derivative through a
redefinition of the gauge fields after an
orthogonal transformation \cite{O'Leary:2011yq}, so that %
\be%
D_{\mu} = \partial_{\mu} - i Q_{\phi}^T G A_{\mu} ,%
\ee %
where $Q_\phi$ is a vector with the $U(1)_Y$ and $U(1)_{B-L}$
charges for the field $\phi$ and $G$ is the gauge coupling matrix%
\be %
G = \left(\begin{array}{cc} g_1 & \tilde{g}\\
0 & g_{B-L} \end{array}\right). %
\ee%
In this basis, one finds%
\bea %
M_Z^2 &=& \frac{1}{4} (g_1^2 + g_2^2) v^2,\\
M_{Z'}^2 &=& g_{B-L}^2 v'^2 + \frac{1}{4} \tilde{g}^2 v^2, %
\eea %
with the following mixing angle between $Z$ and $Z'$:%
\be%
\tan 2 \theta' = \frac{2 \tilde{g} \sqrt{g_1^2 +
g_2^2}}{\tilde{g}^2 + 16 (\frac{v'}{v})^2 g_{B-L}^2 - g_2^2
-g_1^2}.%
\ee%
It is clear that in the limit of $\tilde{g} \rightarrow 0$, which
will be adopted in our analysis, the $Z$ and $Z'$ sectors decouple. In
this case, the $B-L$ gauge symmetry is radiatively broken, as shown
in Ref.\cite{Khalil:2007dr}. The scalar potential
$V(\chi_1,\chi_2)$ is given by %
\be %
V(\chi_1,\chi_2)= \frac{1}{2} g^2_{B-L}(\vert
\chi_2\vert^2 - \vert \chi_1\vert^2)^2 + \mu^2_1 \vert
\chi_1\vert^2 + \mu^2_2\vert \chi_2\vert^2 - \mu^2_3(\chi_1 \chi_2
+
h.c),%
\ee %
where
\be%
\mu^2_i = m_0^2 + \mu^{\prime^2}  (i=1,2),
\hspace{2cm} \mu^2_3 = - B \mu^\prime~. \label{mp12} %
\ee %
The stability of the potential (boundedness from below) implies
\begin{equation}
\mu_1^2 + \mu_2^2 > 2 \vert \mu_3^2\vert ~. %
\label{stab}
\end{equation}
Also, to avoid a vanishing minimum of the scalar potential, the following constraint is
required
\begin{equation}
\mu_1^2 \mu_2^2 < \mu_3^4~ . \label{minimiz}
\end{equation}
At the TeV scale these conditions can be satisfied simultaneously, thanks to
the running of the scalar masses $m_{\chi_1}^2$ and
$m_{\chi_2}^2$ starting from the Grand Unification Theory (GUT) scale. The minimisation of
$V(\chi_1,\chi_2)$ at the TeV scale leads to the following conditions:%
\bea %
v^{\prime^2} = (v^{\prime^2}_1 + v^{\prime^2}_2) &=&
\frac{(\mu^2_1-\mu^2_2) - (\mu^2_1+\mu^2_2)\cos2\beta'}{2
g^2_{B-L} \cos2\beta'},\\ \label{muprime} \sin 2 \beta'
&=& \frac{2
\mu^2_3}{\mu^2_1+\mu^2_2},%
\eea %
where $\langle \chi_1 \rangle = v'_1$ and $\langle \chi_2 \rangle =
v'_2$. The angle $\beta'$ is defined as $\tan \beta' =v'_1/v'_2$. For
a given $M_{Z'}/g_{B-L} > 6$ TeV (as dictated by experimental constraints), the
minimisation condition (\ref{muprime}) can be used to determine
the SUSY parameter
$\mu^{\prime^2}$, up to a sign.  One finds %
\be %
\mu^{\prime^2}= \frac{m_{\chi_2}^2 - m_{\chi_1}^2 \tan^2 \theta }{\tan^2\theta-1} - \frac{1}{4}M_{Z'}^2 . %
\ee%

{We now consider the neutrino/sneutrino sector. After the TeV
scale $B-L$ breaking, the neutrino mass matrix is given by
\be %
M_\nu = \mat{0 & m_D\\ m_D^{\dag} & M_N}, %
\ee%
where $m_D=Y_\nu v_2,\ M_N=Y_N v'_1$. The neutrino masses,
obtained by the diagonalization of such a mass matrix,  are given
by%
\be
m_{\nu_\ell} & \simeq & -m_D M_N^{-1}m_D^{\dag},\\
m_{\nu_H} & \simeq & M_N. \ee Therefore, if $M_N \sim {\cal O}(1)$
TeV, the light neutrinos $\nu_\ell$ mass can be of order one eV if
the Yukawa coupling $Y_\nu$ is of order $10^{-6}$. This small
coupling is of order the electron Yukawa coupling, so it is not
quite unnatural.}

With a TeV scale
right-handed sneutrino, the sneutrino mass matrix, for one generation
in the basis ($\nn_L,\nn_L^\ast,\nn_R,\nn_R^\ast$), is given by the following $4\times 4$ Hermitian matrix:%
\begin{equation}
{\cal M}^2 = \mat{M^2_{LL} & M^2_{LR} \\ \left(M^2_{LR}\right)^{\dag} & M^2_{RR}
},\label{sneutrino-matrix}
\end{equation}
where
\begin{eqnarray}
M^2_{LL} & = & \mat{m_{\tilde{L}}^2+m_D^2+\frac{1}{2}M_Z^2\cos 2\beta-\frac{1}{2}M_{Z'}^2\cos 2\beta' & 0 \\
0 & m_{\tilde{L}}^2+m_D^2+\frac{1}{2}M_Z^2\cos
2\beta-\frac{1}{2}M_{Z'}^2\cos 2\beta'},\\\n
M^2_{LR} & = & \mat{m_D(A_\nu-\mu\cot\beta + M_N) & 0\\
0 & m_D(A_\nu-\mu\cot\beta + M_N)},\\\n
M^2_{RR} & = & \mat{M_N^2 + m_{\tilde{N}}^2 + m_D^2+ \frac{1}{2}M_{Z'}^2\cos 2\beta' & M_N(A_N-\mu'\cot\beta')\\
M_N(A_N-\mu'\cot\beta') &
M_N^2+m_{\tilde{N}}^2+m_D^2+\frac{1}{2}M_{Z'}^2\cos 2\beta'}.
\end{eqnarray}
{
Recalling that the soft SUSY breaking masses, $M_{Z'}$, $A$-terms, $\mu'$ and $M_N$ are of order $\sim 1$ TeV, and $m_D\sim{\cal O}(10^{-4})$ GeV, then}
it is clear that the mixing
between left- and right-handed sneutrinos is quite suppressed since it
is proportional to the Yukawa coupling $Y_\nu \lsim {\cal
O}(10^{-6})$. A large mixing between the right-handed sneutrinos and
right-handed antisneutrinos is quite plausible, since it is given in
terms of the Yukawa term $Y_N \sim {\cal O}(1)$. This illustrates that
$\nn_L,\nn_L^\ast$ are mass eigenstates and they have degenerate
masses equal to $\left(M^2_{LL}\right)_{11}$ whereas
$\nn_R,\nn_R^\ast$ are not mass eigenstates and the masses of the
mass eigenstates will be different from each other. Indeed, the
eigenvalues of the matrix $M^2_{RR}$ are given by
\begin{equation}
m^2_{\tilde{\nu}_{R_{1,2}}} = m^2_{\tilde{\nu}_{R}} \mp \Delta
m^2_{\tilde{\nu}_{R}},
\end{equation}
where $m^2_{\tilde{\nu}_{R}} = \frac{1}{2}(m^2_{\tilde{\nu}_{R_1}}
+ m^2_{\tilde{\nu}_{R_2}})$ is the average of heavy right-handed
sneutrino squared-mass, which is given by%
\be%
m^2_{\tilde{\nu}_{R}}=M_N^2+m_{\tilde{N}}^2+m_D^2+\frac{1}{2}M_{Z'}^2\cos
2\beta'.%
\ee%
While $\Delta m^2_{\tilde{\nu}_{R}}$ is the mass-splitting in the
heavy right-handed sneutrinos, which is given by
\be%
\Delta m^2_{\tilde{\nu}_{R}} = M_N \Big\vert A_N-\mu'\cot\beta' \Big\vert. %
\ee%

The mass splitting and mixing between the right-handed sneutrino
$\tilde{\nu}_R$ and right-antisneutrino $\tilde{\nu}_R^*$
are a result of a $\Delta L =2$  lepton number violation, as intimated. Therefore,
it is natural that the mass splitting between the
right-handed (anti)sneutrinos  is given in terms of the right-neutrino mass
$M_N$, which represents the magnitude of lepton number violation.
This should be compared with any mass splitting between
left-(anti)sneutrinos, which is typically characterised by the
size of the neutrino masses. As a consequence, $\tilde{\nu}_R$ and
$\tilde{\nu}_R^*$ are no longer the mass
eigenstates. The latter are given instead by%
\be%
\tilde{\nu}_1 = \frac{1}{\sqrt2} \Big( \tilde{\nu}_R + \tilde{\nu}_R^* \Big),~~~~~~~~~~~~~~\tilde{\nu}_2 = \frac{-i}{\sqrt2} \Big( \tilde{\nu}_R - \tilde{\nu}_R^* \Big).%
\label{combination}
\ee %
Here, we assumed that the SUSY parameters $A_N$ and $\mu'$ are
real. The $\tilde{\nu}_R$ and $\tilde{\nu}_R^*$ mixing is an
analogue to the mixing in $B^0-\bar{B}^0$ and $K^0-\bar{K}^0$ that
are generated by a $\Delta B=2$ and $\Delta S=2$ violation,
respectively.

\section{Constraints on right-handed sneutrino-antisneutrino mixing}
\label{sec:constraints}

{In this section we explore possible constraints that may be
imposed on the right-handed sneutrino mass splitting. One of such
constraints could be due to the neutrinoless double-$\beta$ decay
experimental limit. However, it turns out that right-handed
sneutrinos may enter such a process only at the one-loop level, hence their contribution
is substantially suppressed. Therefore, no stringent limit on the sneutrino mass
splitting can be obtained in this connection.
Another interesting possibility is the contribution of
right-handed sneutrinos to the one-loop radiative correction onto the
light neutrino masses. Such a correction, due to the left-handed sneutrino mass
splitting, has been calculated in
\cite{grossman-haber1}. In this case, the contribution gives a stringent
constraint on the sneutrino mass splitting because the largest correction
emerges via a wino-like chargino, which has a non-suppressed
coupling to the light neutrinos. Further, we will perform here a
similar calculation of the one-loop correction to the light
neutrino masses, but this time due to the right-handed sneutrinos. As will
be seen, this effect is suppressed and places no constraints on
the right-handed sneutrino mass splitting. This is because the chargino in
this case is Higgsino-like, hence its has a suppressed Yukawa coupling
($Y_\nu \lesssim 10^{-6}$).}

\begin{figure}[!t]
\begin{center}
\includegraphics{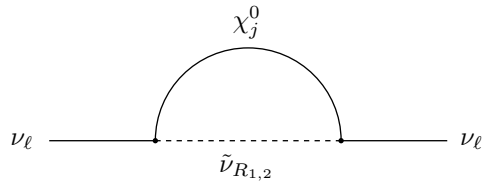}
\caption{One-loop contribution to the neutrino mass due to the
sneutrino mass splitting.} \label{feyn}
\end{center}
\end{figure}

The sneutrino mass splitting may generate one loop contribution to
the neutrino mass \cite{grossman-haber1}. In general, this induces
a stringent constraint on the sneutrino mass splitting. In our
$(B-L)$SSM, the one loop contribution to the neutrino mass is shown in
Fig.~\ref{feyn}, where right-handed sneutrinos and neutral
Higgsinos are running in the loop.
The calculation of this diagram leads to
\begin{equation}
m_\nu^{(1)} = \frac{|Y_\nu|^2 \Delta
m_{\nn_R}}{32\pi^2}\;\sum_{j}|U_{jH}|^2f(x_j,y_j), \label{num1}
\end{equation}
where $x_j=\frac{m_{\nn_{R_1}}^2}{m_{\chi_j}^2},\;y_j=\frac{m_{\nn_{R_2}}^2}{m_{\chi_j}^2}$ and $U_{jH}$ is the
neutralino mixing matrix element which projects the neutralino
onto the Higgsino $\tilde{H}_2$. The loop function $f(x,y)$ is given
by
$$ f(x,y) = \frac{1}{\sqrt y - \sqrt x} \left( \frac{y}{y-1} \ln y - \frac{x}{x-1} \ln x \right).$$
In the limit $m_\nu,\Delta m_{\nn_R} \ll m_{\nn_R}$, the definition of this loop function goes into that of Ref. \cite{grossman-haber1}. The common feature of both of the two functions is that their values does not go more than 1.

As can be seen from eq. (\ref{num1}), the right-handed sneutrino loop contribution to the
neutrino mass is suppressed by the
small Yukawa coupling, $Y_\nu \lsim 10^{-6}$. Assuming that the
neutrino mass, $m_\nu = m_\nu^0 + m_\nu^{(1)}$, is of order ${\cal
O}(10^{-9})$, to satisfy experimental limits. Thus,
$m_\nu^{(1)} \ll 10^{-9}$ GeV, which implies that $\vert Y_\nu
\vert^2 \Delta m_{\tilde{\nu}_R} \ll 10^{-9}$. This bound can
easily be satisfied for any value of $\Delta m_{\tilde{\nu}_R} \sim
{\cal O}(10^{3})$ GeV, since $Y_\nu^2 \ll 10^{-12}$. Therefore,
one can conclude that in the $(B-L)$SSM the light neutrino mass cannot
impose any constraint on the right-handed sneutrino
mass splitting. Accordingly, $x_{\tilde{\nu}_R} = \Delta
m_{\tilde{\nu}_R}/\Gamma_{\tilde{\nu}_R}$, where
$\Gamma_{\tilde{\nu}_R}$ is the average decay rate of
$\tilde{\nu}_R$ and $\Delta m_{\tilde{\nu}_R}= \Delta
m_{\tilde{\nu}_R}^2/2 m_{\tilde{\nu}_R}$ can be quite large. This
implies enough time for right-handed (anti)sneutrino oscillation, that can be
probed by the final state lepton charge. The oscillation of
right-handed sneutrinos into right-antisneutrinos is described by \cite{Huitu-Honkavaara2011}%
\be%
P_{\tilde{\nu}_R \rightarrow \tilde{\nu}_R^*} (t) =
\frac{x_{\tilde{\nu}_R}^2}{2(1+x_{\tilde{\nu}_R}^2)}.%
\ee %
Thus, for $x_{\tilde{\nu}_R} \gg 1$, the right-handed
sneutrino-antisneutrino oscillation probability is $\simeq 1/2$.

\section{LHC results}
\label{sec:results}
%
\begin{figure}[!t]
\begin{center}
\includegraphics[width=9.5cm,height=5.5cm]{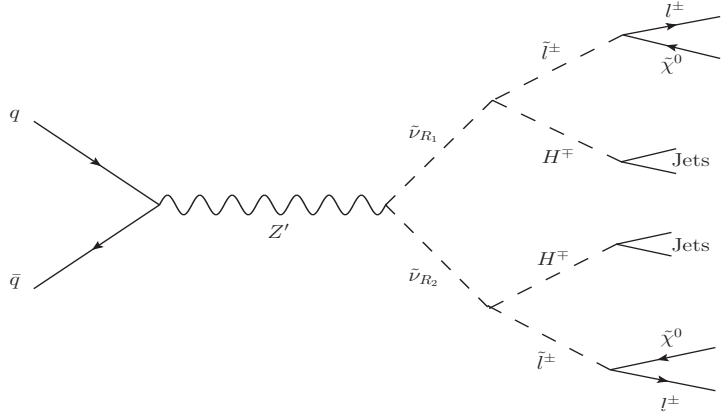}
\caption{Production of right-handed sneutrino pairs at the LHC and its decay
to a same-sign di-lepton pair, missing transverse energy and jets.} \label{likesign2}
\end{center}
\end{figure}
In this section we study the production and decay of
right-handed sneutrinos produced in pairs at the LHC. Also, we will explore the charge
asymmetry of the emerging same-sign di-lepton final state. From the $(B-L)$SSM  Lagrangian, one can show
that the relevant interactions for the right-handed sneutrino
decay are given by%
\bea%
{\cal L}_{int}^{^{\nn_R}} &=& (Y_\nu)_{ij} \bar{l}_i P_R (V_{k2}
\tilde{\chi}^+_k)^\dag (\Gamma_{\nu R})_{\alpha j}
\tilde{\nu}_{R_\alpha} + (Y_{\nu})_{ij}(U_{\rm MNS})_{il} \bar{\nu}_l
P_R (N_{k1}^* \tilde{\chi}_k^0) (\Gamma_{\nu_R})_{j\alpha}
\tilde{\nu}_{R_\alpha}  \nonumber\\
&+& (Y_\nu)_{ij} (M_N)_j \cos \beta \left[(\Gamma_{L_L})_{\beta i} \tilde{l}_\beta H^+ (\Gamma_{\nu R})_{\alpha j}\tilde{\nu}_{R_\alpha} \right].%
\eea%
Here, we assume that the charged leptons are in their physical
basis. The rotational matrices $\Gamma_{L_L}$ and $\Gamma_{\nu_R}$
are defined as $\Gamma_L \equiv (\Gamma_{L_L}, \Gamma_{L_R})$ and
$\Gamma_\nu \equiv (\Gamma_{\nu_L}, \Gamma_{\nu_R})$. Further, the
neutralino mass matrix is diagonalised by a $4 \times 4$ rotation
matrix $N$ and the chargino mass matrix is diagonalised by two
rotation matrices $U,V$. From this equation, it can be easily seen
that if the lightest right-handed sneutrino is heavier than the
lightest slepton, then the former can decay into the latter and a
charged Higgs boson, which in turn decay into SM particles. Since
the coupling right-handed sneutrino-slepton-charged Higgs boson is
proportional to $M_N$, the associated decay rate may not be
suppressed. Accordingly, in this case, the right-handed sneutrino
is no longer a long-lived particle and one should look for a final
state with a same-sign di-lepton pair, missing (transverse) energy
($E_T^{\rm miss}$) and jets, as shown in Fig.~\ref{likesign2}. It
is important to note that both physical right-handed sneutrinos
$\tilde{\nu}_{1,2}$ may decay to $H^+ \tilde{l}^-$ or $H^-
\tilde{l}^+$ with equal probability since, as shown in
(\ref{combination}), they are composed of an equal combination of
$\tilde{\nu}_R$ and $\tilde{\nu}_R^*$.

\begin{figure}[t]
\begin{center}
\epsfig{file=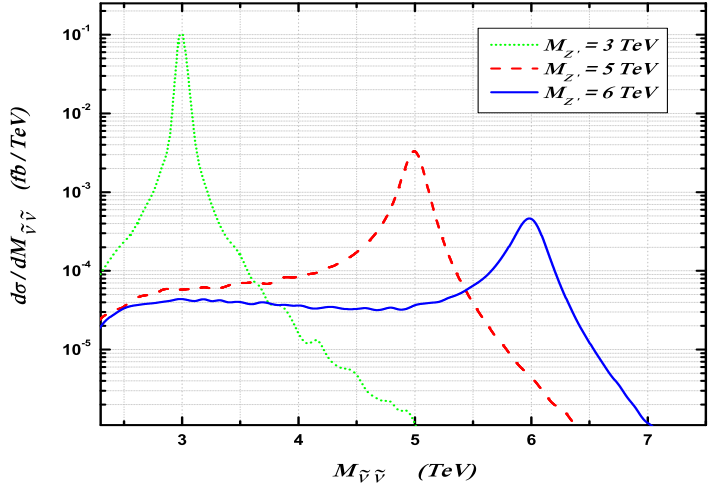, width=10.5cm,height=7.5cm,angle=0}
\end{center}
\vspace{-0.5cm} \caption{The differential distribution in the invariant mass of the sneutrino pair
in the process $pp(q\bar{q})\to Z'\to\tilde{\nu}_{R_1} \tilde{\nu}_{R_2}$ at the LHC for 14 TeV.} %
\label{diff}
\end{figure}

In this case, the total cross section of  such a {\it same-sign di-lepton} ($SS$ for short)
signal at the LHC, which is a most striking signature for right-handed
sneutrino-antisneutrino oscillation, is given by
\bea%
\sigma(q\bar{q}\to Z'\to\tilde{\nu}_{R_1} \tilde{\nu}_{R_2}\to l^-
l^- +E_T^{\rm miss}+\rm{jets})&\simeq&\sigma(q\bar{q}\to
Z'\to\tilde{\nu}_{R_1} \tilde{\nu}_{R_2})\nonumber \\
&\times&
{\rm BR}(\tilde{\nu}_{R_1} \tilde{\nu}_{R_2} \to\tilde{l}^- \tilde{l}^- H^+H^+ \to l^- l^-
+E_T^{\rm miss}+\rm{jets}).%
\eea
The scattering Matrix Element (ME), averaged/summed over initial/final colours and spins, for sneutrino pair production is given by
\be%
|{\cal M}(q\bar{q}\to Z'\to\tilde{\nu}_{R_1} \tilde{\nu}_{R_2})|^2=
C_q (Y^q_{B-L})^2 (Y^{\nu_R}_{B-L})^2
      \frac{8 g_{B-L}^4 s |\vec{k}|^2}
     {(s-M_{Z'}^2)^2+(M_{Z'}\Gamma_{Z'})^2}
     (1-\cos^2\theta),
\ee%
where $C_q=3$ is a colour factor, $Y_{B-L}^q$ and $Y_{B-L}^{\nu_R}$ are the $B-L$ charges of
the quarks and right-handed sneutrinos, respectively,
$\sqrt{s}$ is the Centre-of-Mass (CM) energy of the partonic
collision at the LHC, $\cos\theta$ is the polar angle in the CM frame and
\be
|\vec k|^2=\frac{(s-m_{{\tilde\nu}_{R_1}}^2-m_{{\tilde\nu}_{R_2}}^2)^2
            -(2 m_{{\tilde\nu}_{R_1}} m_{{\tilde\nu}_{R_2}} )^2}
         {4 s}
\ee
is the modulus squared of the three-momentum of either sneutrino in the final state.
(See the last paper in Ref.~\cite{B-L-LHC} for the expressions yielding $\Gamma_{Z'}$.)
 In Fig. \ref{diff} we show the differential  cross
section of right-handed sneutrino pair production at the LHC with $\sqrt s=14$ TeV as CM energy
 for three choices
of $M_{Z'}$, 3, 5 and 6 TeV, and $g_{B-L}$, 1/2, 5/6, 1,
correspondingly, each compliant with the limit from EW Precision Tests
(EWPTs), i.e., $M_{Z'}/g_{B-L}>6$ TeV. The corresponding values of the integrated cross sections
are 11, 1 and 0.3 fb, respectively~\footnote{We have used here the CTEQ6L1 \cite{cteq} Parton Distribution Functions
(PDFs) with factorisation/renormalisation scale set to $\mu=Q=\sqrt{s}$.}. For reference, we have taken
$m_{{\tilde\nu}_{R_1}}=0.8$ TeV and $m_{{\tilde\nu}_{R_2}}=1.2$ TeV. The cross sections are therefore observable with standard luminosity.

%

Note that, if the charged Higgs mass is less than $200$ GeV, then
its dominant decay channel is $H^{\pm}\to \tau^{\pm}\nu_{\tau}$
and accordingly $\textrm{BR}(H^{\pm}\to
\tau^{\pm}\nu_{\tau})\simeq 1$ . However, for $m_{H^{\pm}} > 200$
GeV, the decay channel $H^{\pm}\to t b$ becomes dominant and one
finds that $\textrm{BR}(H^{\pm}\to l^{\pm} \nu_l)\simeq {\cal
O}(0.1)$ \cite{Djouadi2}. Also, if the charged slepton is assumed
to be the second lightest SUSY particle and the lightest
neutralino is the lightest one, then the
$\textrm{BR}(\tilde{l}^{\pm}\to l^{\pm} \tilde{\chi}^0)\simeq {\cal O}
(1)$.

\begin{figure}[t]
\begin{center}
\epsfig{file=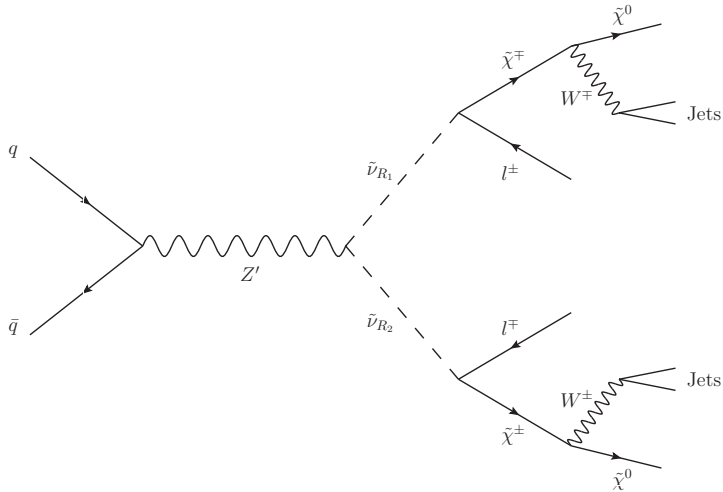,width=9.5cm,height=6.5cm,angle=0}
\end{center}
\vspace{-0.5cm} \caption{Production of right-handed sneutrino pairs at the LHC and its decay
to an opposite-sign di-lepton pair, missing transverse energy and jets.} \label{lpm}
\end{figure}

In contrast, if the mass of the right-handed sneutrino is smaller than
the mass of the slepton, then the only available decay channels for the
right-handed sneutrino are: $\tilde{\nu}_{R_{1,2}} \to l^\pm \tilde{\chi}^\mp$ or
$\tilde{\nu}_{R_{1,2}} \to \nu_{L} \tilde{\chi}^0$\cite{KhalilLFV}. In the former case,
the chargino may decay to $W^\mp$ and the lightest neutralino.
Therefore, an {\it opposite-sign di-lepton} ($OS$ for short) pair,
missing transverse energy and jets, is a possible signal as shown in Fig. \ref{lpm}. It is worth mentioning
that due to the oscillation between the right-handed sneutrino and
antisneutrino, which is reflected in the mass difference between
$\tilde{\nu}_{R_1}$ and $\tilde{\nu}_{R_2}$, it is possible for
$\tilde{\nu}_{R_1}$ to decay to $l^-$ whilst $\tilde{\nu}_{R_2}$
decays to $l^+$.
The difference between {\it SS} and {\it OS} outgoing di-leptons implies what is
known as the {\it lepton charge asymmetry}, which can be measured at
the LHC, providing a smoking gun signal for right-handed sneutrino
oscillation. The lepton charge asymmetry is defined as \cite{Huitu-Honkavaara2009}%
\bea A^{{\rm asym}}=\frac{\sigma(SS) -
\sigma(OS)}{\sigma(SS)+\sigma(OS)} =
\frac{\sigma(l^-l^-\tilde{\chi}^+\tilde{\chi}^+)-\sigma(l^+l^-\tilde{\chi}^+\tilde{\chi}^-)}{\sigma(l^-l^-\tilde{\chi}^+\tilde{\chi}^+)+\sigma(l^+l^-\tilde{\chi}^+\tilde{\chi}^-)}\
,\eea
where the $SS$ cross section s is obtained as
\begin{eqnarray}
\sigma(SS) = \sigma(q\bar{q} \rightarrow Z' \rightarrow
\tilde{\nu}_{R_1} \tilde{\nu}_{R_2})\;{\rm BR}(\tilde{\nu}_{R_1}
\rightarrow l^+ \tilde{\chi}^-)\;{\rm BR}(\tilde{\nu}_{R_2} \rightarrow
l^+ \tilde{\chi}^-),
\end{eqnarray}
and the $OS$ cross section  is given by
\begin{eqnarray}
\sigma(OS) = \sigma(q\bar{q} \rightarrow Z' \rightarrow
\tilde{\nu}_{R_1} \tilde{\nu}_{R_2})\;{\rm BR}(\tilde{\nu}_{R_1}
\rightarrow l^+ \tilde{\chi}^-)\;{\rm BR}(\tilde{\nu}_{R_2} \rightarrow
l^- \tilde{\chi}^+).
\end{eqnarray}
Here, we assume that the  primary leptons produced in the (anti)sneutrino decays are always distinguishable from
those that may emerge from the decays of the charginos. Furthermore, owing to SM background, we note that the $OS$ signal will be seen with more difficulty than the $SS$ one. However, the significant amount of missing energy it presents can be exploited to remove contamination from $Z+$jet, $W^+W^-+$jet events as well as from pure QCD noise (in presence of leptonic decays of hadrons).

Recall that the physical right-handed sneutrino states are defined
as
\begin{eqnarray}
\tilde{\nu}_{R_1} & = & \tilde{\nu}_{R} \cos\alpha + \tilde{\nu}_{R}^* \sin\alpha,\\
\tilde{\nu}_{R_2} & = & -\tilde{\nu}_{R} \sin\alpha +
\tilde{\nu}_{R}^* \cos\alpha,
\end{eqnarray}
where the mixing angle $\alpha$ depends on the sneutrino and
antisneutrino mass difference. Now, for large $\Delta
m_{\tilde{\nu}_R}$ one obtains $\alpha=\frac{\pi}{4}$. {This can be seen also from the fact that $\alpha$ is expressible in terms of
the entries of the mass matrix $M_{RR}^2$,
\be
\cos 2\alpha = \frac{\left(M_{RR}\right)_{11}-\left(M_{RR}\right)_{22}}{\sqrt{\left(\left(M_{RR}\right)_{11}-\left(M_{RR}\right)_{22}\right)^2+4\left(M_{RR}\right)_{12}^2}}=0,
\ee
which correspond to maximal mixing between $\tilde{\nu}_{R}$ and
$\tilde{\nu}_{R}^*$.} In these conditions, one has
BR$(\tilde{\nu}_{R_1} \rightarrow\; l^+ \tilde{\chi}^-) =$
BR$(\tilde{\nu}_{R_2} \rightarrow\; l^- \tilde{\chi}^+) = {\cal O}(1)$
(and charge conjugates).
Therefore, the cross sections $\sigma(SS)$ and
$\sigma(OS)$ are identical.
Explicitly, the lepton charge asymmetry
$A^{\rm asym}$ can be written as
\begin{eqnarray}
A^{\rm asym} & = & \frac{{\rm BR}(\tilde{\nu}_{R_2}\; \rightarrow\; l^+
\tilde{\chi}^-)-{\rm BR}(\tilde{\nu}_{R_2}\; \rightarrow\; l^-
\tilde{\chi}^+)}{{\rm BR}(\tilde{\nu}_{R_2}\; \rightarrow\; l^+
\tilde{\chi}^-)+{\rm BR}(\tilde{\nu}_{R_2}\; \rightarrow\; l^- \tilde{\chi}^+)}\n & =
& \frac{\cos^2\alpha-\sin^2\alpha}{\cos^2\alpha+\sin^2\alpha}=
\cos 2\alpha.\label{asym}
\end{eqnarray}


It is therefore clear that, if there is no oscillation, the lepton charge asymmetry will
be given by $A^{\rm asym}=-1$, while with maximal oscillation the
asymmetry is given by $A^{\rm asym}=0$. An {\it effective lepton charge asymmetry}
is commonly introduced to overcome the misleading  effect associated to the
fact that the maximal mixing condition corresponds the value $A^{\rm asym}=0$. This is
defined as%
\begin{eqnarray}
A_{\textrm{eff}} = \frac{A^{\rm asym} + 1}{2},
\label{effA}
\end{eqnarray}
Therefore, the effective lepton charge asymmetry
associated to the decay of right-handed sneutrinos is given by
\begin{eqnarray}
A_{\textrm{eff}} = \frac{1}{2}.
\end{eqnarray}

This result is slightly larger than the charge asymmetry
associated with left-handed sneutrinos decays, studied  in
Ref. \cite{Huitu-Honkavaara2011}. In that scenario, the mass
difference between right-handed sneutrino and antisneutrino is
quite small, however a small sneutrino decay width ($\sim
10^{-14}$) was assumed, which implies that $x_{\tilde{\nu}} \sim
{\cal O}(1)$ and hence significant oscillation can still take
place. Our result of a large effective lepton charge asymmetry gives the hope that
the right-handed sneutrino-antisneutrino can be easily probed at
the LHC. However, this clearly depends upon the error associated with the asymmetry observables.

\begin{figure}[t]
\begin{center}
\epsfig{file=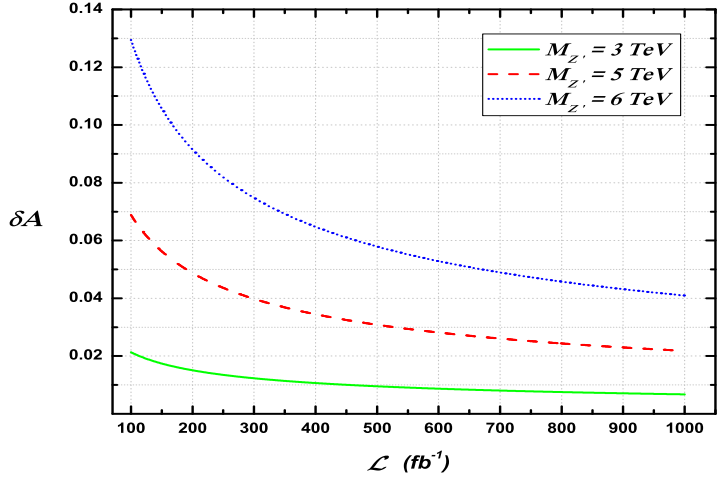, width=10.5cm,height=7.5cm,angle=0}
\end{center}
\vspace{-0.5cm} \caption{The statistical error on the lepton charge asymmetry in eq.~(\ref{deltaA})
versus the luminosity at the 14 TeV LHC.}%
\label{error}
\end{figure}

For a  given luminosity ${\cal L}$, such that $N_{SS}={\cal L}\sigma(SS)$ and $N_{OS}={\cal L}\sigma(OS)$ are the
event rates for the $SS$ and $OS$ final states,
the statistical error of the
predicted lepton charge asymmetry is obtained as \cite{book}%
\be%
\delta A^{\rm asym} = 2 \sqrt{\frac{N_{SS} N_{OS}}{(N_{SS} + N_{OS})^3}}.%
\label{deltaA}
\ee%
 In Fig. \ref{error}, such a quantity is presented as a function of the LHC luminosity, at 14 TeV, for our aforementioned choices of
$M_{Z'}$ and $g_{B-L}$. As it can
be seen, this error is already small enough at 100 fb$^{-1}$ to enable a statistically significant extraction of $A_{\rm eff}$ and
it becomes less that $1\%$ for ${\cal
L} \sim 1000$ fb$^{-1}$. In fact, note that the corresponding errors for the
effective lepton charge asymmetry are given by 1/2 those
 reported in this figure, as the coefficient 1/2 in front of $A^{\rm asym}$ in eq.~(\ref{effA})
is a precisely known real-valued constant. Note that for smaller values of the LHC energy, 7 and 8 TeV, the errors in the effective lepton charge asymmetry are enhanced significantly due to the associated suppression for cross section and luminosity. In particular, for $\sqrt{s}=7(8)$ TeV, one finds that $\sigma=0.1593(0.5016), 0.0003(0.0012), 8.59\times 10^{-5}(0.0003)$ fb for $M_{Z'}=3,5,6$ TeV, and in this case $\delta A_{\textnormal{eff}}$ would be of order $0.41(0.22),9.96(4.61),17.1(9.56)$, assuming ${\cal L}=5(20)$ fb$^{-1}$. Hence, it is clearly only the high energy and luminosity setup of the CERN machine that will enable one to fully probe sneutrino oscillations.

We should of course mention that our discussion should eventually be subject to validation following detailed phenomenological analyses in particular accounting for background effects, which were not dealt with here, also in presence of the decays of the chargino pair. This is however beyond the scope of this paper that, on the one hand, made the general point that the oscillation phenomenology emerging in the extended SUSY model considered here is just as quantitatively significant as in other SUSY scenarios studied in previous literature and that, on the other hand, implicitly relies upon the fact that successful signal-to-background analyses similar to those carried out in those contexts can be repeated in the present one.

Finally, we should like to mention the possibility that the two right-handed
sneutrinos decay into light SM-like neutrinos and lightest
neutralinos. This decay channel, if it took place, would be an
invisible channel, since both light neutrinos and lightest
neutralinos would be escaping the detector. This signal may provide therefore a
robust signature for $B-L$ sneutrino oscillations, say, a mono-jet or single photon, essentially free from
SM background. However, we leave the pursuit of the
consequent phenomenology to a separate publication \cite{prep}.

\section{Summary and conclusions}
\label{sec:summa} In conclusion, we have proven that right-handed
sneutrino-antisneutrino oscillations, emerging in the $(B-L)$SSM
in presence of a type I seesaw mechanism of light neutrino mass
generation, are testable at the LHC. In fact, after taking into
account the constraints imposed on the mass splitting between
heavy right-handed sneutrino and the corresponding antisneutrino
by the experimental limits set on the light neutrino masses, we
have shown that pair production of such right-handed sneutrinos
decaying into leptons and charginos generates a cross section
which is promptly accessible at 14 TeV and an effective lepton
charge asymmetry that can be resolved already after 100 fb$^{-1}$
of luminosity, both of which offer an efficient means to resolve
the aforementioned oscillation phenomenon. {Finally, it is worth
mentioning that the signature of sneutrino-antisneutrino
oscillations can also be obtained from other possible extensions of the
MSSM, that lead to $\Delta L=2$ violation, like the MSSM with $R$-parity
violation or with Higgs triplets or else a SUSY Left-Right
model. In this respect, our analysis is quite relevant and it is
not limited to the $B-L$ extension of MSSM that we have adopted
here.}

\section*{Acknowledgments}
S.K. thanks The Leverhulme Trust (London, UK) for financial support in the form of a Visiting Professorship to the University
of Southampton. S.M. is financed in part through the NExT Institute.


\end{document}